\documentclass[apj]{emulateapj}

\shorttitle{RR Lyrae stars in Tucana}
\shortauthors{Bernard et al.}

\begin{document}

\title{The ACS LCID Project: RR Lyrae stars as tracers of old population
    gradients\\ in the isolated dwarf spheroidal galaxy Tucana\altaffilmark{1}}

\author{Edouard J. Bernard,\altaffilmark{2} Carme Gallart,\altaffilmark{2}
    Matteo Monelli,\altaffilmark{2} Antonio Aparicio,\altaffilmark{2,3}
    Santi Cassisi,\altaffilmark{4} \\ Evan D. Skillman,\altaffilmark{5}
    Peter B. Stetson,\altaffilmark{6} Andrew A. Cole,\altaffilmark{7}
    Igor Drozdovsky,\altaffilmark{2,8} Sebastian L. Hidalgo,\altaffilmark{5} \\
    Mario Mateo,\altaffilmark{9} Eline Tolstoy\altaffilmark{10}}

\altaffiltext{1}{Based on observations made with the NASA/ESA Hubble Space
 Telescope, obtained at the Space Telescope Science Institute, which is
 operated by the Association of Universities for Research in Astronomy, Inc.,
 under NASA contract NAS 5-26555. These observations are associated with
 program \#10505.}
\altaffiltext{2}{Instituto de Astrof\'{i}sica de Canarias,
    La Laguna, Tenerife, Spain; ebernard@iac.es, monelli@iac.es, carme@iac.es,
    antapaj@iac.es, dio@iac.es.}
\altaffiltext{3}{Departamento de Astrof\'{i}sica, Universidad de La Laguna,
    Tenerife, Spain.}
\altaffiltext{4}{INAF-Osservatorio Astronomico di Collurania, Teramo,
    Italy; cassisi@oa-teramo.inaf.it.}
\altaffiltext{5}{Department of Astronomy, University of Minnesota,
    Minneapolis,
    USA; slhidalgo@astro.umn.edu, skillman@astro.umn.edu.}
\altaffiltext{6}{Dominion Astrophysical Observatory, Herzberg Institute of
    Astrophysics, National Research Council, Victoria, Canada;
    peter.stetson@nrc-cnrc.gc.ca.}
\altaffiltext{7}{School of Mathematics \& Physics, University of Tasmania,
    Hobart, Tasmania, Australia; andrew.cole@utas.edu.au.}
\altaffiltext{8}{Astronomical Institute, St. Petersburg State University,
    St. Petersburg, Russia.}
\altaffiltext{9}{Department of Astronomy, University of Michigan,
    Ann Arbor, USA; mmateo@umich.edu.}
\altaffiltext{10}{Kapteyn Astronomical Institute, University of Groningen,
    Groningen, Netherlands; etolstoy@astro.rug.nl.}

\begin{abstract}

 We present a study of the radial distribution of RR Lyrae variables, which
 present a range of photometric and pulsational properties, in the dwarf
 spheroidal galaxy Tucana. We find that the fainter RR Lyrae stars, having a
 shorter period, are more centrally concentrated than the more luminous, longer
 period RR Lyrae variables.
 Through comparison with the predictions of theoretical models of stellar
 evolution and stellar pulsation, we interpret the fainter RR Lyrae stars as a
 more metal-rich subsample. In addition, we show that they must be older than
 about 10~Gyr. Therefore, the metallicity gradient must have appeared very
 early on in the history of this galaxy.

\end{abstract}


\section{Introduction}

 The oldest stellar populations in galaxies are of fundamental importance
 because they can constrain the epoch of onset of star formation in the
 Universe. Unfortunately, the oldest main-sequence turn-offs (MSTO), which
 provide the most straightforward characterization of the old population, can
 only be reached in a handful of Milky Way satellite galaxies using current
 ground-based observing facilities. An alternative approach is the
 characterization of the RR~Lyrae star populations. The fact that the
 properties of the RR Lyrae stars reflect the properties of the old population
 is crucial because the horizontal-branch (HB) stars are about three
 magnitudes brighter than their MSTO counterpart.

 For example, it has long been suspected that the mean magnitude of the
 RR Lyrae variables, M$_V$(RR), is a function of [Fe/H], in the sense that more
 metal-poor RR Lyrae stars are more luminous \citep{san58}. The correlation
 between M$_V$(RR) and [Fe/H] was later confirmed by stellar evolution models,
 but with the additional constraint that the morphology of the HB, i.e., the
 evolutionary status of the stellar population, must be taken into account
 \citep[e.g.,][]{cas04}. As the population gets older, the mass of the stars
 reaching the HB decreases, giving a bluer HB. In this case, the stars
 crossing the instability strip (IS) are therefore already evolving off the
 zero-age horizontal branch (ZAHB) and are more luminous than younger stars of
 the same metallicity.

 In addition, observations of Galactic globular clusters suggested that the
 lower their metallicity, the longer the mean period of their RR Lyrae stars.
 This led to the various period-metallicity relations found in the literature
 \citep[see, e.g.,][]{cat92,san93,sar06}, or the more elaborate
 period-metallicity-amplitude relations \citep[e.g.,][]{alc00,san04},
 providing a convenient way of estimating the metallicity of a stellar system
 based on distance- and reddening-free observables (namely, mean period
 $\langle P_{ab} \rangle$ and visual amplitude A$_V$).

 We note, though, that \citet{cle99} have argued that the period-amplitude
 relation for RR$ab$ stars is not a function of metal abundance, but is
 instead related to the evolutionary status of the RR Lyrae stars. In the
 metal-poor, Oosterhoff type II clusters, the (blue) HB stars would have
 started to evolve away from the ZAHB, crossing the IS at higher luminosities,
 therefore following a different PA relation from the ZAHB RR Lyrae stars.

 However, theoretical \citep{bon07} and observational
 \citep{weh99,kal00,lay00,ole01,bor01} evidence suggest that, while there is
 a clear correlation of Oosterhoff type with metallicity, there is an
 additional dependence of the period at a given amplitude upon metallicity,
 with the more metal-rich stars having shorter periods.

 An analysis of the variation of these observables as a function of the
 position within a galaxy can highlight the eventual radial trends across the
 studied galaxy. This, in turn, provides important clues to the formation
 mechanisms and the star formation history of the host galaxy. Hence, by
 providing information about the properties of the underlying population,
 variable star research procures a way to study the histories of these
 galaxies, independent and complementary to the CMD analysis.

 In this {\it Letter} we report, for the first time, the detection of stellar
 population gradients in a dwarf galaxy based solely on the periods and
 luminosities of its RR Lyrae stars. This was made possible by the large
 number of RR Lyrae variables discovered in the isolated Local Group dSph
 Tucana, based on {\it Hubble Space Telescope} ({\it HST}) ACS data covering
 the whole body of the galaxy, together with the high quality of the light
 curves. We present supporting evidence from the different evolutionary phases
 present in the CMD.

\section{Observations and Data Reduction}

 The present analysis is based on observations obtained with the Advanced
 Camera for Surveys onboard the {\it HST} as part of the major program
 LCID\footnotemark[11] aiming at deriving detailed star formation histories for
 a sample of Local Group isolated dwarf galaxies (C. Gallart et al. 2008, in
 preparation).
 As the goal of these observations was to reach the oldest main sequence
 turn-offs with good signal-to-noise on the final stacked images, 32 orbits
 were devoted to the observations of Tucana. These were collected over about
 five consecutive days between 2006 April 25 and 30. As each orbit was devoted
 to one exposure in the F475W band and one in F814W, the observing sequence
 consisted of alternating $\sim$1000 seconds exposures in F475W and F814W for
 an optimal sampling of the light curves.
 The DAOPHOT/ALLFRAME suite of programs \citep{ste94} was used to obtain the
 instrumental photometry of the stars on the individual, non-drizzled images.
 A detailed description of the observations, data reduction and calibration
 will be given in a forthcoming paper (M. Monelli et al. 2008, in preparation).
 To search for variable stars, the Welch-Stetson variability index
 \citep{wel93} was used. From the list of candidates, 358 stars were identified
 as bona fide RR Lyrae stars, for which the period was searched through Fourier
 analysis following the prescription of \citet{hor86}. We identified 216
 RR$ab$, 82 RR$c$, and 60 RR$d$ based on their period and light-curve shape.
 Further details about the variable star content of Tucana and another
 isolated dSph, Cetus, will be presented in a future paper (E. Bernard et al.
 2008, in preparation).

 \footnotetext[11]{Local Cosmology from Isolated Dwarfs,
 http://www.iac.es/project/LCID/}

\section{The RR Lyrae Population and its Spatial Gradients}

\begin{figure}
\epsscale{1.2} 
\plotone{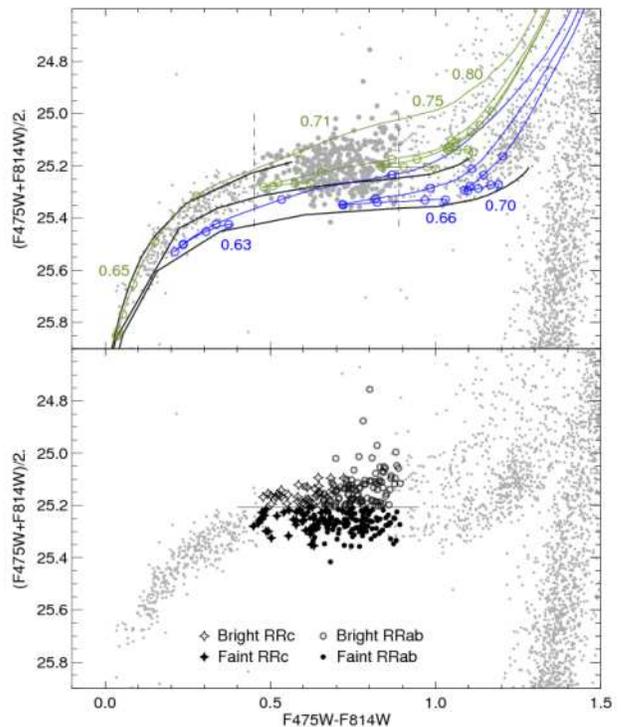}
\figcaption{Zoom-in on the HB of Tucana. {\it Top}: The RR Lyrae stars are
 overplotted with larger dots. The solid black lines represent, from top to
 bottom, the Z=0.0001, 0.0003, and 0.001 ZAHB. The green and blue lines are
 evolutionary tracks with Z=0.0003 and Z=0.001, respectively, and masses as
 labeled. Overplotted on them, the circles indicate intervals of 10 Myr.
 ZAHB loci and tracks are scaled-solar models from the BaSTI library
 \citep{pie04}, plotted assuming A$_g$=0.122 and A$_I$=0.061 from NASA/IRSA
 (http://irsa.ipac.caltech.edu/applications/DUST/) and (m-M)$_0$=24.84
 (C. Gallart et al. 2008, in preparation).
 The vertical dashed lines roughly delimit the IS.
 {\it Bottom}: Different symbols have been used for each type and subsample of
 RR Lyrae variables, as labeled in the figure. For the sake of clarity, RR$d$
 are not represented. The horizontal line shows the mean intensity-weighted
 magnitude of all the RR Lyrae stars.
\label{fig:1}}
\end{figure}

\begin{figure}
\epsscale{1.2} 
\plotone{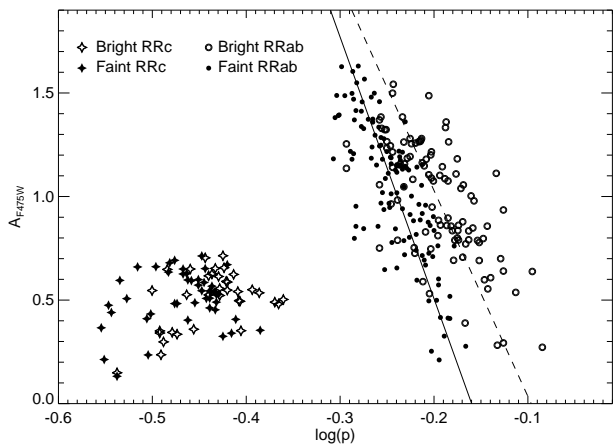}
\figcaption{Period-amplitude diagram for the RR Lyrae stars in Tucana. Symbols
 are as in the bottom panel of Fig.~\ref{fig:1}. Linear fits to each sample of
 RR$ab$ are also represented.
\label{fig:2}}
\end{figure}

 Figure~\ref{fig:1} ({\it top}) shows the CMD of Tucana centered on the HB.
 The $(F475W+F814W)/2 \sim V$ filter combination was chosen for the ordinate
 axis so that the HB appears approximately horizontal. The
 RR Lyrae stars are overplotted with larger dots using their intensity-weighted
 magnitudes. One can see that Tucana harbors a rather complex HB, which is well
 populated on both sides of the IS. The red side also presents a small gap in
 magnitude, suggesting the combination of two HB of different luminosities.
 This is supported by the unusual width in luminosity of the HB inside the
 IS ($\sim$0.3 mag at F475W-F814W$ = $0.7, excluding the few bright and faint
 outliers). As the evolutionary tracks show, in Fig.~\ref{fig:1} ({\it top}),
 a stellar population at a given metallicity does not span more than $\sim$0.15
 mag in brightness within the instability strip, even taking into account the
 evolution off the ZAHB.

\begin{figure}
\epsscale{1.2} 
\plotone{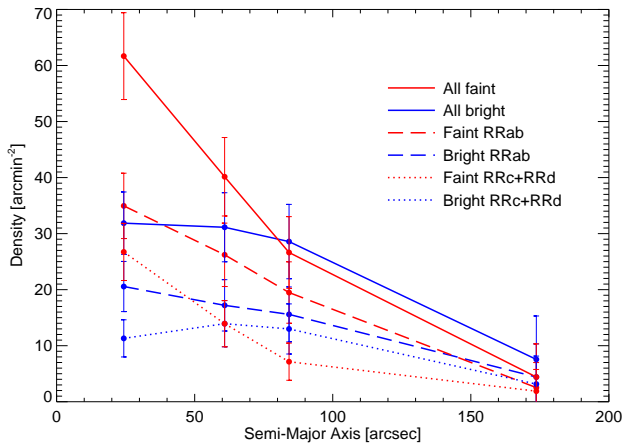}
\figcaption{Radial distribution of the different subsamples of RR Lyrae stars
 in Tucana, showing how the fainter RR Lyrae stars are systematically more
 concentrated in the center of the galaxy than the brighter variables.
 The error bars correspond to Poisson statistics.
\label{fig:3}}
\end{figure}

 We arbitrarily used the mean magnitude of the variables at (F475W+F814W)/2.
 $\sim$ 25.2, shown with the horizontal line in the bottom panel of
 Fig.~\ref{fig:1}, to split the RR$ab$, RR$c$ and RR$d$ variables into bright
 and faint subsamples, each type of variable having approximately the same
 number of stars in each subsample. Different symbols have been used for each
 type and subsample. For the sake of clarity, RR$d$ variables are not
 represented.

 The RR Lyrae stars are plotted on the period-amplitude (PA) diagram in
 Fig.~\ref{fig:2}. The symbols are the same as in the bottom panel of
 Fig.~\ref{fig:1}. Linear fits to each sample of RR$ab$ are also represented.
 We can clearly identify two distinct RR$ab$ sequences, characterized by
 different mean periods and dispersions around the fit. Based on their period
 distribution, we find a Kolmogorov-Smirnov (KS) discrepancy of
 D$_{KS}$=0.4397, implying a probability P$_{KS}<$$10^{-9}$\% that the bright
 and faint RR$ab$ subsamples were drawn from the same parent population.
 As predicted by nonlinear pulsation models \citep{bon97b}, the mean period is
 a function of luminosity, in the sense that the more luminous variables tend
 to have a larger period.
 The period difference also shows up in the RR$c$ subtype: the ``bell-shape"
 distribution \citep{bon97b} of the brighter RR$c$ is shifted toward longer
 period with respect to the fainter RR$c$. The difference in mean period is
 $\sim$0.02 days (P$_{KS}<$$0.34$\%). The mean primary period of double-mode
 RR Lyrae stars present a similar shift between the bright and faint
 subsamples (P$_{KS}<$$2 \times 10^{-5}$\%).

 Interestingly, the faint and bright subsamples of RR$ab$ variables have mean
 periods of 0.574 and 0.640 respectively, which is very reminiscent of the
 mean periods of the Oosteroff types I and II \citep[0.55 and 0.64,][]{oos39}.
 This is also the case for the subsamples of RR$c$ variables, having mean
 periods of 0.343 and 0.366, versus 0.32 and 0.37 for the OoI and OoII
 globular clusters of the Milky Way. On the other hand, the whole sets of
 RR$ab$ and RR$c$ have mean periods of 0.604 and 0.353, respectively. These
 numbers would lead to the classification of Tucana as an Oosterhoff
 intermediate (Oo-Int), as is the case for most dSph \citep{cat04}. In this
 particular case, however, there is the possibility that the Oo-Int status
 could be attributed to the mixture of OoI and OoII populations, as hinted by
 the two roughly parallel sequences in the PA diagram of the RRab in
 Fig.~\ref{fig:2}.

\begin{figure}
\epsscale{1.25} 
\plotone{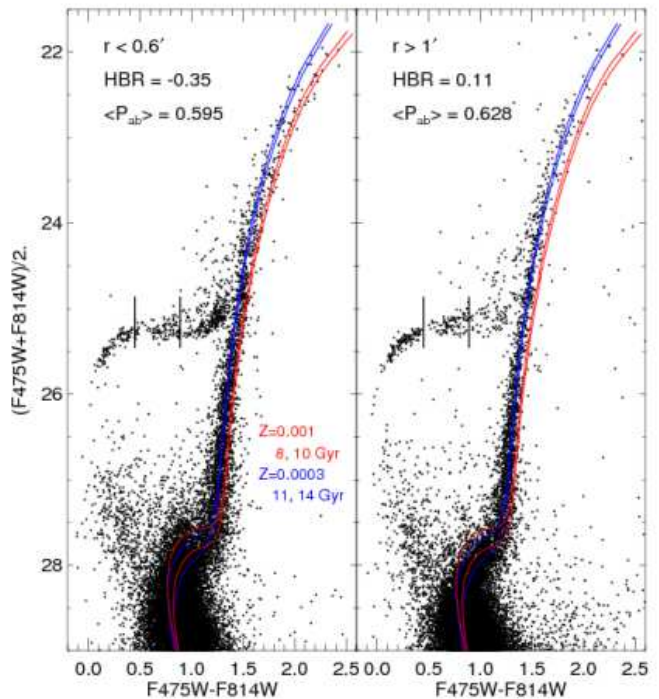}
\figcaption{Color-magnitude diagrams of the inner and outer regions of Tucana,
 containing the same number of stars. The vertical lines delimit the IS.
 Isochrones from the BaSTI library \citep{pie04}, with ages and metallicities
 indicated in the figure, have been superimposed to highlight the gradients.
 Note the difference in the upper RGB, HB/red clump morphology and upper
 main-sequence.
\label{fig:4}}
\end{figure}

 The presence of different populations in a galaxy, whether due to the details
 of its star formation history or to the accretion of an external stellar
 system, generally leads to gradients in the observable properties of its
 stars. Figure~\ref{fig:3} presents the radial profile for each subsample of
 RR Lyrae stars. The radii were chosen so that each concentric region contains
 the same number of variables.
 It shows that, for each type of RR Lyrae stars, the fainter variables are
 systematically more concentrated near the center of the galaxy, while the
 brighter RR Lyrae stars are spatially extended.
 The difference in spatial distribution is very significant as the radii
 containing half of the bright and faint RR Lyrae stars are 1.33$\arcmin$ and
 1.09$\arcmin$, respectively, and a KS test gives a probability of $\la$0.001
 for the bright and faint RR Lyrae stars to have the same radial distribution.
 The combination of different intrinsic properties of the individual stars
 with the different spatial distribution supports the hypothesis that they
 represent separate populations.

\begin{figure}
\epsscale{1.2} 
\plotone{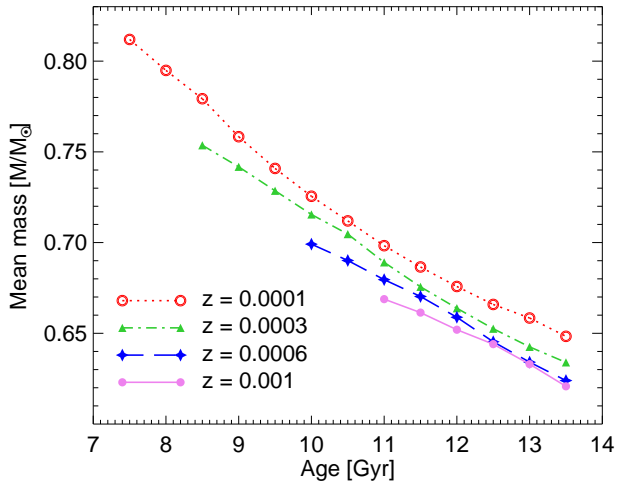}
\figcaption{Mean mass as a function of age and metallicity for artificial stars
 {\it within the IS} of Tucana, from IAC-STAR simulations using the BaSTI
 library with $\eta$=0.4 \citep{pie04} and no dispersion in mass loss at a
 given age.
\label{fig:5}}
\end{figure}

\section{Discussion}

 The difference in the mean period of the RR$ab$ stars of each sample could be
 attributed to both a difference in metallicity (from the period-metallicity
 correlation) or a difference in the evolutionary status of the individual
 stars, the stars in the redward evolution off the ZAHB having a longer period
 \citep[their Fig.~16-17]{bon97b}. However, the top panel of Fig.~\ref{fig:1}
 shows that evolution off the ZAHB alone cannot account for the range of
 luminosity spanned by the RR Lyrae stars. Hence, it is necessary to invoke a
 range of metallicity to reproduce the distribution of stars within the
 IS. The hypothesis of a bimodal metallicity distribution is strengthened by
 the double ``bell-shape" of the RR$c$ in the PA diagram \citep{bon97a} and the
 presence of two RGB bumps separated by $\sim$0.2 mag in F814W (C. Gallart et
 al. 2008, in preparation).
 We use the luminosity of the theoretical ZAHB in Fig.~\ref{fig:1} to estimate
 this range. A firm upper limit for the metallicity of the old population is
 given by the ZAHB at Z=0.001. While the Z=0.0001 ZAHB seems to be slightly
 too luminous, the Z=0.0003 tracks cannot produce the brightest RR Lyrae stars
 in our observations. We suggest Z$\sim$0.0002 as an approximate lower limit
 to the metallicity of Tucana.

 Although the range of metallicity is relatively wide for a dwarf spheroidal
 galaxy, it is supported by the color of the RGB of the inner and outer regions
 of Tucana presented in Fig.~\ref{fig:4}.
 The isochrones show that the higher metallicity population visible in the
 central region, with Z$\sim$0.001, is absent in the outer region, while a
 metallicity slightly lower than Z=0.0003 is needed to explain the blue side of
 the sub-HB RGB.
 Note, also, the difference in the morphology of the HB. The faint RR Lyrae
 variables and the red HB stars show up only in the center of the galaxy,
 while the brighter stars of the IS and the blue HB stars are present
 everywhere.
 Consequently, the HB ratio (HBR\footnotemark[12]) increases by about 0.5 over
 the radius sampled with our data, similar to what was observed in the Sculptor
 dSph \citep[][see also \citealt{hur99}]{maj99} for which spectroscopy
 indicates the presence of a metallicity gradient \citep{tol04} and a
 metallicity spread \citep{cle05} within the old ($\ga$10~Gyr) HB population.

 \footnotetext[12]{HBR = (B-R)/(B+V+R), where B, V, and R are the numbers of
 stars to the blue, within, and to the red of the IS \citep{lee90}.}

 Our metallicity range supports the results of \citet{har01}, who found a
 bimodal [Fe/H] distribution in Tucana from the RGB stars over their observed
 {\it HST} WFPC2 field. From isochrones in the HB morphology-metallicity plane,
 they explain the gradient in HB morphology as a `pure metallicity effect'.
 However, artificial CMDs produced by IAC-STAR
 \citep[][http://iac-star.iac.es/iac-star/]{apa04} show that to obtain a red HB
 component such as the one observed in the center of Tucana, stars having a
 total mass in the range $\sim$0.75-0.80 M$_\Sun$ are needed (see, also, the
 top panel of Fig.~\ref{fig:1}).
 Assuming the common value for the mass loss parameter $\eta$=0.40 with the
 BaSTI library \citep{pie04}, these masses correspond to stars younger than
 about 9~Gyr old, quite independently of the metallicity between Z=0.0001 and
 0.001. The same models give an age of $\sim$13 Gyr for the bluest HB stars at
 Z=0.0001 which are present over the whole galaxy.

 The morphology of the MSTO in Fig.~\ref{fig:4} reflects this range of age: one
 can see that the CMD of the central region harbors a younger population than
 the outer region. Fitting isochrones to the subgiant branch gives a population
 age ranging from about 11 to 14~Gyr old in the outer part of the galaxy, while
 it appears to be as young as $\sim$8~Gyr in the center.
 Note that crowding tends to broaden the features in a CMD and may be partly
 decreasing the estimated age of the youngest MSTO stars in the center of the
 galaxy.
 Nonetheless, a preliminary SFH of the inner and outer regions of
 Tucana---which takes into account the effects of crowding---do show that there
 is a slightly younger population in the center.

 However, the youngest stars {\it within} the IS are slightly older than the
 red HB stars.
 Figure~\ref{fig:5} illustrates the results of our artificial HB computations:
 it represents the mean mass as a function of age and metallicity for
 artificial stars within the instability strip of Tucana, and indicates that
 the faintest, highest metallicity (Z$\ga$0.0006) variables of Tucana must be
 $\ga$10~Gyr old.
 {\it Therefore, under the reasonable assumption that chemical enrichment
 follows age in star forming galaxies, the presence of gradients in the
 RR~Lyrae populations shows that these metallicity gradients appeared very
 early on in the history of this galaxy}.

\acknowledgments

{\it Facility:} \facility{HST (ACS)}

 We would like to thank G. Bono for reading the manuscript and useful comments.
 Support for this work was provided by a Marie Curie Early Stage Research
 Training Fellowship of the European Community's Sixth Framework Programme
 under contract number MEST-CT-2004-504604, the IAC (grant P3/94), the Spanish
 Education and Science Ministry (grant AYA2004-06343), and NASA through grant
 GO-10505 from the Space Telescope Science Institute, which is operated by
 AURA, Inc., under NASA contract NAS5-26555.
 This work has made use of the IAC-STAR Synthetic CMD computation code.
 IAC-STAR is supported and maintained by the computer division of the Instituto
 de Astrof\'isica de Canarias.
 This research has made use of the NASA/IPAC Infrared Science Archive, which is
 operated by the Jet Propulsion Laboratory, California Institute of Technology,
 under contract with the National Aeronautics and Space Administration.

\end{document}